\documentclass[cits]{PoS}

\usepackage{sidecap}
\usepackage{epsfig}

\def\NPA{{\em Nucl. Phys.}   {\bf A}}
\def\NPB{{\em Nucl. Phys.}   {\bf B}}
\def\PLB{{\em Phys. Lett.}   {\bf B}}
\def\PRL{{\em Phys. Rev. Lett.}}

\def\PRD{{\em Phys. Rev.}    {\bf D}}
\def\ZPA{{\em Z. Phys.}      {\bf A}}
\def\ZPC{{\em Z. Phys.}      {\bf C}}
\def\EJA{{\em Eur. Phys. J.} {\bf A}}
\def\EJC{{\em Eur. Phys. J.} {\bf C}}

\def\Wgp{W_{\gamma p}}
\def\Wgpi{W_{\gamma\pi}}
\def\gprho{\gamma p \to \rho^0 n \pi^+}

\def\sgp{\sigma_{\gamma p}}
\def\sgpi{\sigma_{\gamma\pi}}
\def\ptn{p_{T,n}}
\def\ptr{p_{T,\rho}}
\def\d2sxp{{\rm d^2}\sgp/{\rm d}x_L{\rm d}\ptn^2}
\def\fluxpi{f_{\pi/p}}
\def\fluxpia{\fluxpi(x_L,t)}
\def\fluxg{f_{\gamma/e}}
\def\Pom{I\!\!P}

\def\aP{\alpha_{\Pom}}

\def\api{\alpha_{\pi}(t)}
\def\Rege{\aP(0)-2\api}

\def\gp{\gamma p}
\def\gpi{\gamma\pi}
\def\dsig{{\rm d}\sigma/{\rm d}}
\title{Exclusive $\rho^0$ Meson Photoproduction with a Leading Neutron at HERA}

\ShortTitle{Photoproduction of $\rho^0$ Meson with a Leading Neutron}

\author{\speaker{Sergey Levonian}%
        \thanks{On behalf of the H1 Collaboration.}\\
        DESY, Notkestra{\ss}e 85, 22607 Hamburg, Germany\\
        E-mail: \email{levonian@mail.desy.de}}

\abstract{A first measurement is presented of exclusive photoproduction
    of $\rho^0$ mesons associated with leading neutrons at HERA.
    The data were taken with the H1 detector in the years $2006$ and $2007$
    at a centre-of-mass energy of $\sqrt{s}=319$ GeV
    and correspond to an integrated luminosity of $1.16$ pb$^{-1}$.
    The $\rho^0$ mesons with transverse momenta $p_T<1$ GeV
    are reconstructed from their decays to charged pions,
    while leading neutrons carrying a large fraction
    of the incoming proton momentum, $x_L>0.35$, are detected
    in the Forward Neutron Calorimeter.
    The phase space of the measurement is defined by the photon virtuality
    $Q^2 < 2$ GeV$^2$, the total energy of the photon-proton system
    $20 < \Wgp < 100$ GeV and the polar angle of the leading neutron
    $\theta_n < 0.75$ mrad.
    The cross section of the reaction $\gamma p \to \rho^0 n \pi^+$
    is measured as a function of several variables.
    The data are interpreted in terms of a double peripheral process,
    involving pion exchange at the proton vertex followed by elastic
    photoproduction of a $\rho^0$ meson on the virtual pion.
    In the framework of one-pion-exchange dominance
    the elastic cross section of photon-pion  scattering,
    $\sigma^{\rm el}(\gamma\pi^+ \to \rho^0\pi^+)$, is extracted.
    The value of this cross section indicates significant
    absorptive corrections for the exclusive reaction $\gprho$.}

\FullConference{XXIV International Workshop on Deep-Inelastic Scattering and Related Subjects \\
                11-15 April, 2016 \\
                DESY Hamburg, Germany}

\begin{document}

\section{Introduction}
\begin{figure}[hht]
\center
 \epsfig{file=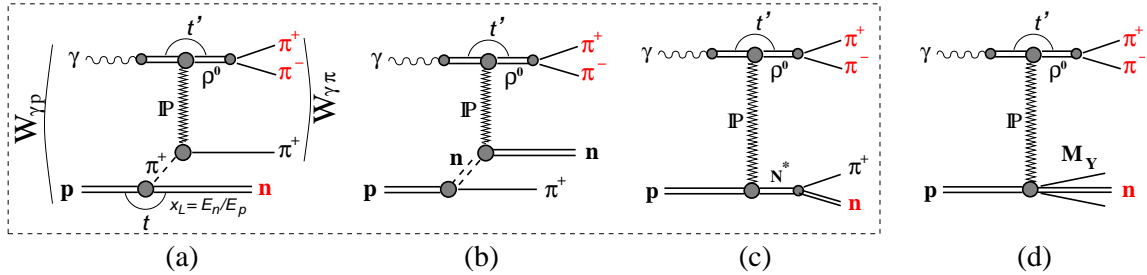,width=\textwidth}
\caption{Generic diagrams for processes contributing to exclusive
         photoproduction of $\rho^0$ mesons associated with leading neutrons 
         at HERA. The signal corresponds to the Drell-Hiida-Deck model graphs
         for the pion exchange (a), neutron exchange (b) and direct pole (c).
         Diffractive scattering in which a neutron may be produced as a part 
         of the proton dissociation system, $M_Y$, contributes as background (d).
         The $N^*$ in (c) denotes both resonant (via $N^+$) and possible 
         non-resonant $n+\pi^+$ production.}
\label{fig:FD}
\end{figure}

The aim of the analysis is to measure exclusive $\rho^0$ production on virtual pion
in the photoproduction regime at HERA and to extract for the first time experimentally
elastic $\gamma\pi$ cross section. In the Regge framework the events of such class are
explained by the diagram shown in Fig.~\ref{fig:FD}a which involves an exchange of two
Regge trajectories in the process $2 \to 3$, known as {\em Double Peripheral Process} (DPP),
or Double-Regge-pole exchange reaction~\cite{DPP}.  
It has been demonstrated in similar reactions at fixed target experiments that
two further diagrams (Fig.~\ref{fig:FD}b, \ref{fig:FD}c) have to be included 
in addition to the pion exchange (Fig.~\ref{fig:FD}a), 
as well as interference between them~\cite{DHD}.
However at small values of four-momentum transfer squared $t$ the contributions 
from graphs \ref{fig:FD}b and \ref{fig:FD}c largely cancel 
and the one-pion-exchange (OPE) term dominates the cross section. 

Events of this type are modelled by the two-step Monte Carlo generator POMPYT~\cite{Pompyt}
in which the virtual pion is produced at the proton vertex according to one of the available
pion flux parametrisations. This pion then scatters elastically on the photon from the
electron beam, thus producing vector meson ($\rho^0$ in our case).
Diffractive dissociation of the proton into a system $Y$ (Fig.~\ref{fig:FD}d) represents
a background to DPP, which contributes due to the limited detector acceptance in the forward
($p$-beam direction) region.
This background is modelled by the DIFFVM generator~\cite{DiffVM} and is statistically
subtracted from the data.


\section{Analysis Outline and Main Results}
%
Using VDM~\cite{VMD} flux $\fluxg(y,Q^2)$ to convert $ep$ cross section into $\gamma p$ one, 
and one-pion-exchange approximation~\cite{OPE} to decompose photon-proton cross section into
a pion flux convoluted with a photon-pion cross section one obtains for the reaction
of interest $e + p \rightarrow e + \rho^0 + n + \pi^+$
\begin{equation}
  \frac{{\rm d}^2\sigma_{ep}}{{\rm d}y {\rm d}Q^2} = \fluxg(y,Q^2) \sgp(\Wgp(y)); \hspace*{0.5cm}
  \frac{{\rm d}^2\sgp(\Wgp,x_L,t)}{{\rm d}x_L{\rm d}t} = \fluxpia \, \sgpi(\Wgpi) ,
  \label{eq:xsec}
\end{equation}
with the generic form of the pion flux factor~\cite{pi_flux}:
\begin{equation}
   \fluxpia = \frac{1}{2\pi} \frac{g_{p\pi n}^2}{4\pi}
                   (1-x_L)^{\Rege}\frac{-t}{(m_{\pi}^2-t)^2} F^2(t,x_L) .
  \label{eq:piflux}
\end{equation}
Here $\aP(0)$ is the Pomeron intercept,
$\api=\alpha^{\prime}_{\pi}(t-m_{\pi}^2)$ is the pion trajectory,
$g_{p\pi n}^2/4\pi$ is the $p\pi n$ coupling constant
known from phenomenological analysis of low energy data,
and $F(t,x_L)$ is a form factor accounting for off mass-shell corrections
and normalised to unity at the pion pole, $F(m_{\pi}^2,x_L)=1$.  
For exact definition of kinematic variables in eq.~(\ref{eq:xsec}-\ref{eq:piflux}) see~\cite{H1paper}.

The analysis is based on $\sim 6600$ events, containing only two charged pions from $\rho^0$ decay
and a leading neutron with energy $E_n>120$ GeV, and nothing else above noise level in the detector.
The sample corresponds to an integrated luminosity of $1.16$ pb$^{-1}$, collected
by a special minimum bias track trigger in the years 2006-2007 at $\sqrt{s_{ep}}=319$ GeV.
Further details of the analysis can be found in~\cite{H1paper}.

\begin{figure}[hhh]
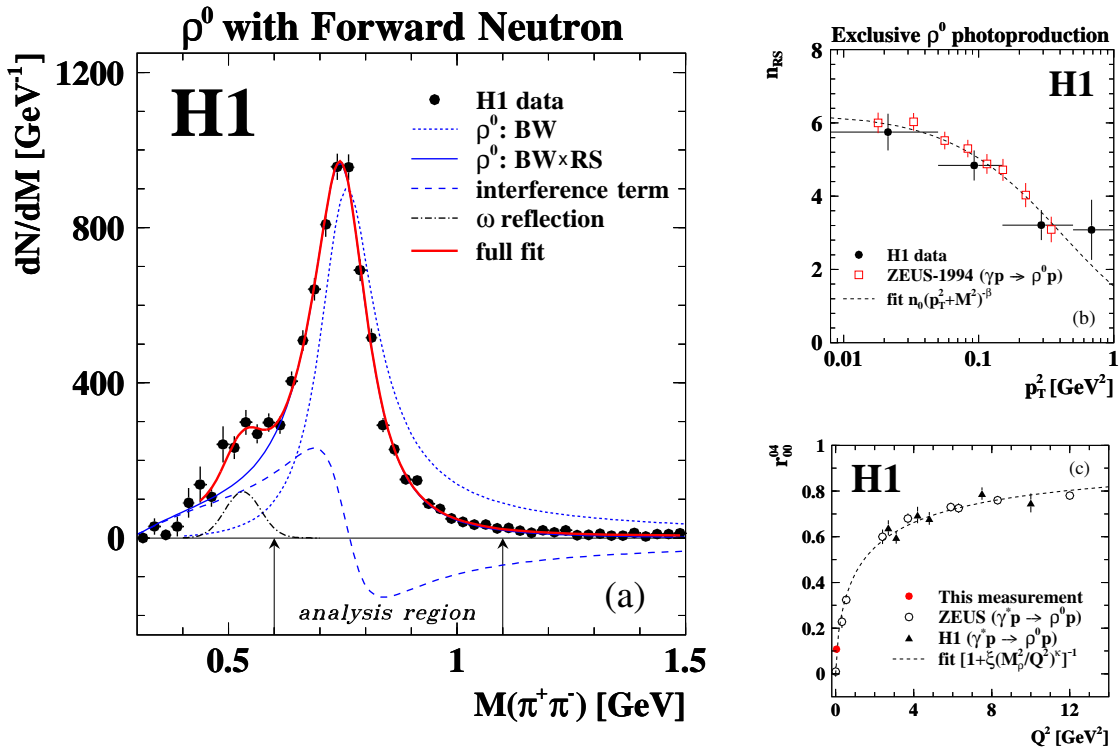

\center
 \setlength{\unitlength}{1cm}
 \begin{picture}(15,9.8)(0,0)
   \put(-.1,0.0){\epsfig{file=fig_02a.eps,width=0.65\textwidth}}
   \put(10.,4.7){\epsfig{file=fig_02b.eps,width=0.34\textwidth}}
   \put(10.,0.0){\epsfig{file=fig_02c.eps,width=0.34\textwidth}}
 \end{picture}
\caption{The $\rho^0$ meson properties:
         (a) Mass distribution of the $\pi^+\pi^-$ system 
         with $p_T^2<1.0$ GeV$^2$.
         The data points are corrected for the detector efficiency.
         The curves represent different components contributing
         to the measured distribution and the Breit-Wigner resonant part
         extracted from the fit to the data.
         The analysis region $0.6<M_{\pi^+\pi^-}<1.1$ GeV is indicated
         by vertical arrows.
         (b) Ross-Stodolsky skewing parameter, $n_{RS}$, as a function
         of $p_T^2$ of the $\pi^+\pi^-$ system.
         The values measured in this analysis
         are compared to previously obtained results
         for elastic photoproduction of $\rho^0$ mesons,
         $\gamma p \to \rho^0 p$, by the ZEUS Collaboration.
         (c) Spin-density matrix element, $r_{00}^{04}$, as a function
         of $Q^2$ for diffractive $\rho^0$ photo- and
         electro-production.
         The curves on figures (b,c) represent the results of the fits
         discussed in~\cite{H1paper}.}
\label{fig:mass}
\end{figure}

The effective mass distribution for two charged pions with $p_t>200$ MeV each
and within the central detector range $20^o<\theta<160^o$ is shown in Fig.~\ref{fig:mass}a.
The distribution is corrected for the mass dependent detector efficiency. The sample is very 
clean with only $1.5\%$ background contamination in the analysis region.
The $p_T$ dependent distortion of the Breit-Wigner line shape is shown in Fig.~\ref{fig:mass}b.
In Fig.~\ref{fig:mass}c the spin-density matrix element $r_{00}^{04}$ as measured in this experiment 
is compared to the values obtained at HERA for different $Q^2$ ranges.


After all selections described in~\cite{H1paper} the remaining proton dissociative background 
fraction in the sample is estimated as $0.34\pm 0.05$. This value is verified by normalising
DIFFVM prediction to the orthogonal, background dominated sample, in which additional 
activity originating from $M_Y$ system (see Fig.\ref{fig:FD}d) in the forward region was required.

The $\gp$ cross section integrated in the domain
$0.35\!<\! x_L\!<\! 0.95$ and $\ptr\!<\! 1$ GeV
and averaged over the energy range $20 \!<\! \Wgp \!<\! 100$ GeV
is determined for two intervals of leading neutron transverse momentum,
corresponding to the full angular acceptance $\theta_n < 0.75$ mrad 
and the ``OPE-safe'' region respectively:
\begin{equation}
  \sigma (\gamma p \to \rho^0 n \pi^+ ) = (310 \pm 6_{\rm stat} \pm 45_{\rm sys})~ {\rm nb}
         \hspace*{0.8cm} {\rm for} \hspace*{0.6cm} \ptn<x_L \cdot 0.69 {\rm ~GeV}
  \label{eq:sgp1}
\end{equation}
\begin{equation}
  \sigma (\gamma p \to \rho^0 n \pi^+ ) = (130 \pm 3_{\rm stat} \pm 19_{\rm sys})~ {\rm nb}
         \hspace*{0.8cm} {\rm for} \hspace*{1.2cm} \ptn<0.2 {\rm ~GeV}.
  \label{eq:sgp2}
\end{equation}

\begin{figure}[hhh]
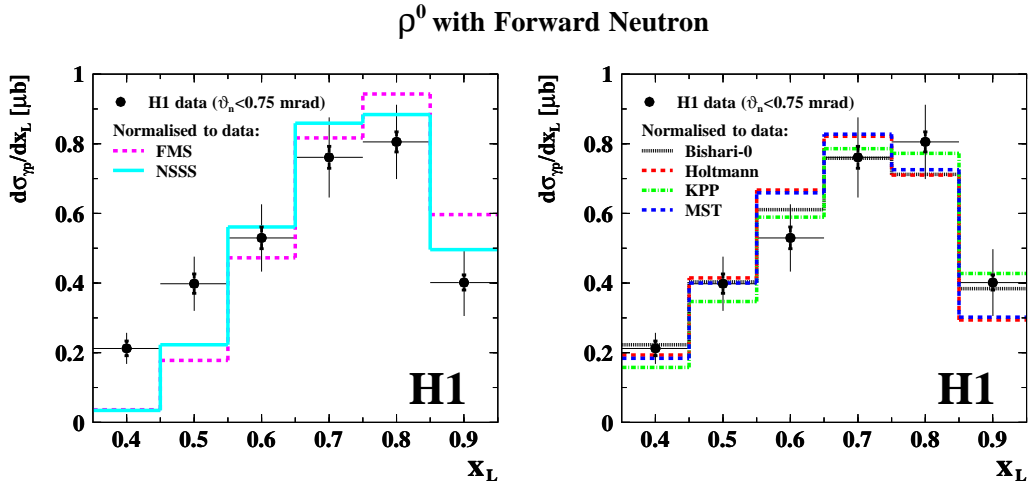

\center
 \setlength{\unitlength}{1cm}
  \begin{picture}(15,0.4)(0,0)
    \put(3.3,0.0){\epsfig{file=fig_tit.eps,width=0.6\textwidth}}
  \end{picture} 
 \epsfig{file=fig_03a.eps,width=0.45\textwidth}
 \epsfig{file=fig_03b.eps,width=0.45\textwidth}
\caption{Differential cross section d$\sigma_{\gamma p}/{\rm d}x_L$ 
         in the range $20<\Wgp<100$ GeV compared
         to the predictions based on different versions of pion fluxes~\cite{pi_flux}.
         The data points are shown with statistical (inner error bars) and total
         (outer error bars) uncertainties, excluding an overall normalisation
         error of $4.4\%$. All predictions are normalised to the data.}
\label{fig:dsdxl}
\end{figure}

The differential cross section d$\sigma_{\gamma p}/{\rm d}x_L$ is shown in Fig.\ref{fig:dsdxl}.
Predictions from several $\pi$-flux models~\cite{pi_flux} are confronted with these data.
Only a shape comparison is possible, since the $\gamma\pi$ cross section is not known.
However, some models can be excluded even on the basis of this shape comparison.


\begin{figure}[hhh]
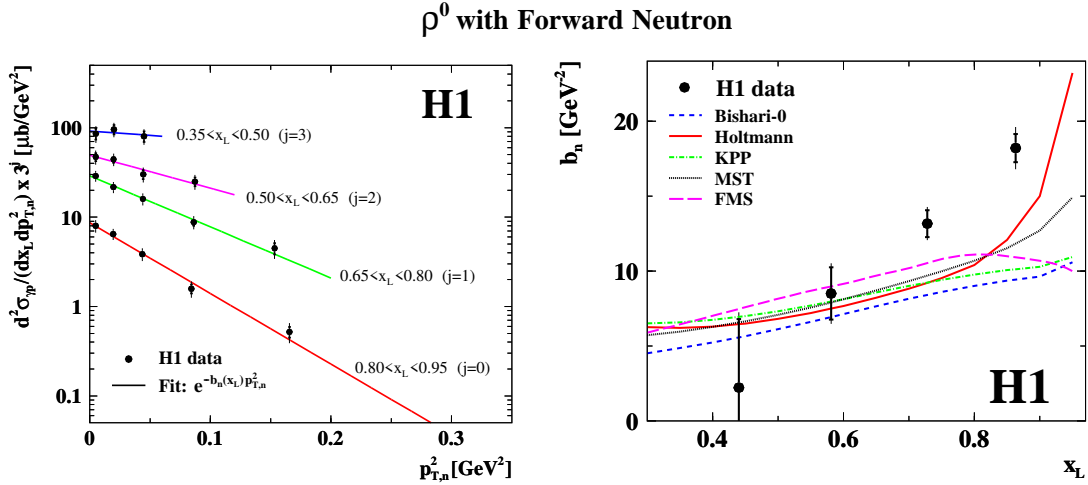

\center
 \setlength{\unitlength}{1cm}
  \begin{picture}(15,0.8)(0,0)
    \put(3.0,0.2){\epsfig{file=fig_tit.eps,width=0.6\textwidth}}
  \end{picture} 
 \epsfig{file=fig_04a.eps,width=0.48\textwidth,clip= }
 \epsfig{file=fig_04b.eps,width=0.51\textwidth,clip= }
\caption{(left) Double differential cross section $\d2sxp$ of neutrons
         in the range $20<\Wgp<100$ GeV
         fitted with single exponential functions.
         The cross sections in different $x_L$ bins $j$ are scaled by
         the factor $3^j$ for better visibility.
         The data points are shown with statistical (inner error bars)
         and total (outer error bars) uncertainties excluding
         an overall normalisation error of $4.4\%$.
         (right) The exponential slopes fitted through the $p_T^2$ dependence 
         of the leading neutrons as a function of $x_L$.
         The inner error bars represent statistical errors and the outer
         error bars are statistical and systematic errors added in quadrature.
         The data points are compared to the expectations of
         several parametrisations of the pion flux within the OPE model.}
\label{fig:bn_xl}
\end{figure}

Additional constraints on the pion flux models could be provided by
the dependence on $t$ (or $\ptn^2$) of the leading neutron.
The double differential cross section $\d2sxp$ is measured,
and the results are presented in Fig.~\ref{fig:bn_xl} (left).
The bins are chosen such, that the data are not affected by the polar angle cut.
The cross sections are fitted by a single exponential function
$e^{-b_n(x_L)\ptn^2}$ in each $x_L$ bin
and the results are presented in Fig.~\ref{fig:bn_xl} (right). 
The measured $b$-slopes are compared to
those obtained from several pion flux parametrisations.
Despite of the large experimental uncertainties none of the models
is able to reproduce the data.
This observation supports phenomenological expectation~\cite{abs_corr} 
of large absorptive corrections which modify the $t$ dependence of the amplitude, 
leading to an increase of the effective $b$-slope at large $x_L$
as compared to the pure OPE model without absorption.


\begin{figure}[hhh]
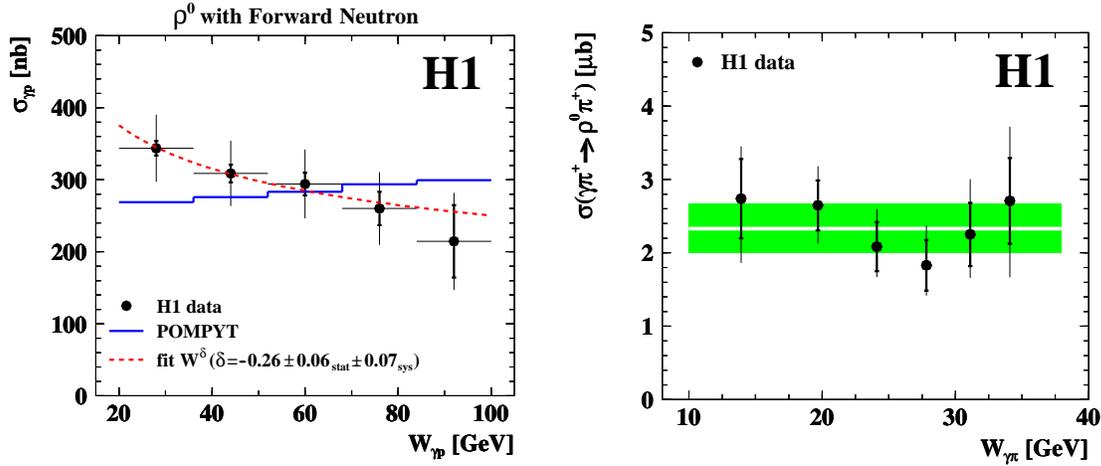

\center
 \epsfig{file=fig_05a.eps,width=0.49\textwidth}
 \epsfig{file=fig_05b.eps,width=0.49\textwidth}
 \setlength{\unitlength}{1cm}
\caption{(left) Cross section of the reaction $\gprho$ as
         a function of $\Wgp$ compared to the prediction 
         from POMPYT MC program, which is normalised to the data.
         The dashed curve represents the Regge motivated fit
         $\sigma \propto W^{\delta}$
         with $\delta = -0.26 \pm 0.06_{\rm stat}\pm 0.07_{\rm sys}$.
         The data points are shown with statistical (inner error bars) and
         total uncertainties (outer error bars) excluding an overall
         normalisation error of $4.4\%$.
         (right) Elastic cross section,
         $\sgpi^{\rm el} \equiv \sigma ({\gamma\pi^+} \to \rho^0\pi^+)$,
         extracted in the one-pion-exchange approximation
         as a function of the photon-pion energy, $\Wgpi$.
         The inner error bars represent the total experimental uncertainty
         and the outer error bars are experimental and model uncertainties
         added in quadrature, where the model error is due to
         pion flux uncertainties.
         The dark shaded band represents the average value for the full $\Wgpi$
         range.}

\label{fig:wgp}
\end{figure}

Fig.~\ref{fig:wgp} (left) shows the energy dependence of exclusive $\rho^0$ production
with a leading neutron, \\ $\sigma_{\gamma p \to \rho^0n\pi^+}(W_{\gamma p})$.
Regge motivated fit $\sigma_{\gamma p} \propto W^{\delta}$ yields a value of
$\delta = -0.26 \pm 0.06_{\rm stat} \pm 0.07_{\rm sys}$.
POMPYT MC predicts different trend, typical for Pomeron exchange only.


The pion flux models compatible with the data in shape of the $x_L$
distribution are used to extract the photon-pion cross sections
from $\dsig x_L$ in the OPE approximation.
The result is presented in in Fig.~\ref{fig:wgp} (right).
As a central value the Holtmann flux  is used,
and the largest difference to the other three predictions
({\em Bishari-0, KPP, MST})~\cite{pi_flux} provides an estimate 
of the model uncertainty which is $\sim\!19\%$ on average.
From the total $\gp$ cross section in equation~(\ref{eq:sgp2})
and using the pion flux~(\ref{eq:piflux}) 
integrated in $x_L$ and $\ptn$, $\Gamma_{\pi} = 0.056$,
the cross section of elastic photoproduction of $\rho^0$ on a pion target
is determined at an average energy $\langle \Wgpi \rangle \simeq 24$ GeV:
\begin{equation} 
      \sigma (\gamma \pi^+\to \rho^0\pi^+) =
      (2.33\pm0.34 (\rm exp) ^{+0.47}_{-0.40} (\rm model))~\mu\rm b,
 \label{eq:sgp3}
\end{equation}
where the model error is due to the uncertainty in the pion flux integral
obtained for the different flux parametrisations compatible with our data.
This value leads to the ratio
$r_{\rm el} = \sigma_{\rm el}^{\gamma\pi}/\sigma_{\rm el}^{\gamma p} = 0.25 \pm 0.06$.
A similar ratio, but for the total cross sections at $\langle W \rangle=107$ GeV,
has been estimated by the ZEUS collaboration as
$r_{\rm tot} = \sigma_{\rm tot}^{\gamma\pi}/\sigma_{\rm tot}^{\gamma p}
= 0.32 \pm 0.03$~\cite{ZEUS_pi2p}.
Both ratios are significantly smaller than
their respective expectations, based on simple considerations.
For $r_{\rm tot}$, a value of $2/3$ is predicted by the additive quark model, while
      $r_{\rm el} = (\frac{b_{\gp}}{b_{\gpi}}) \cdot
      (\sigma_{\rm tot}^{\gpi}/\sigma_{\rm tot}^{\gp})^2 = 0.57 \pm 0.03$
can be deduced by combining the optical theorem, the eikonal approach 
and the world data on $pp, \pi^+p$ and $\gp$ elastic scattering.
Such a suppression of the cross section is usually attributed to rescattering, or 
absorptive corrections~\cite{abs_corr}, which are essential for leading neutron production.
For the exclusive reaction $\gprho$ studied here this would imply
an absorption factor of $K_{abs} = 0.44 \pm 0.11$.

\begin{SCfigure}[0.7][hhb]
 \epsfig{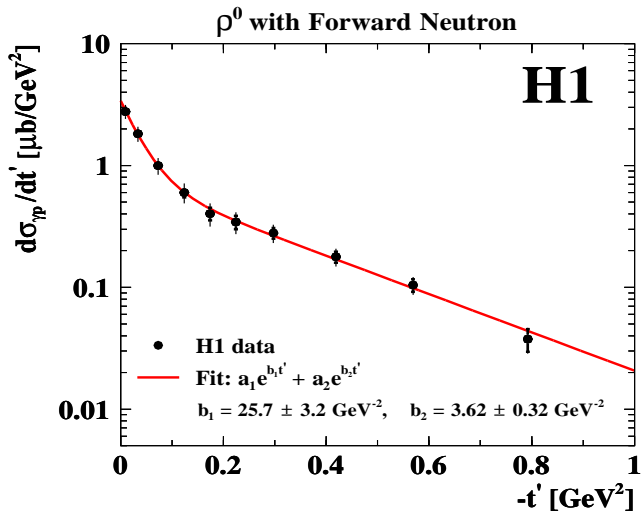}
 \setlength{\unitlength}{1cm}
\caption{Differential cross section d$\sigma_{\gamma p}/{\rm d}t^{\prime}$
         of $\rho^0$ mesons fitted with the sum of two exponential functions.
         The inner error bars represent statistical and uncorrelated
         systematic uncertainties added in quadrature
         and the outer error bars are total uncertainties,
         excluding an overall normalisation error of $4.4\%$.
         The values of slopes are characteristic for double peripheral
         processes~\cite{DPP} in which an exchange of
         two Regge trajectories is involved.}
\label{fig:pt2}
\end{SCfigure}

Finally, the cross section as a function of the four-momentum
transfer squared of the $\rho^0$ meson, $t^{\prime}$,
is presented in Fig.~\ref{fig:pt2}.
It exhibits the very pronounced feature of a strongly changing slope
between the low-$t^{\prime}$ and the high-$t^{\prime}$ regions,
a feature known to be a characteristic for DPP reactions~\cite{DPP}.
In a geometric picture, the large value of $b_1$ suggests that
for a significant part of the data $\rho^0$ mesons are produced
at large impact parameter values of order
$\langle r^2\rangle = 2b_1\!\cdot\!(\hbar c)^2 \simeq 2 {\rm fm}^2
                      \approx (1.6 R_{\rm p})^2$.
In other words, photons find pions in a cloud which extends far beyond
the proton radius.
The small value of $b_2$ corresponds to a target size of $\sim\!0.5$ fm.


\section{Summary}
Photoproduction cross section for exclusive $\rho^0$ production associated with
leading neutron is studied for the first time at HERA. 
Single and double differential $\gp$ cross sections are measured.
The differential cross section d$\sigma/{\rm d}t^{\prime}$ shows the behaviour
typical for exclusive double peripheral exchange processes.
The elastic photon-pion cross section, 
$\sigma(\gamma\pi^+ \to \rho^0\pi^+)$, at $\langle W \rangle = 24$ GeV
is extracted in the OPE approximation.
The estimated cross section ratio for the elastic photoproduction
of $\rho^0$ mesons on the pion and on the proton,
$r_{\rm el} = \sigma_{\rm el}^{\gamma\pi}/\sigma_{\rm el}^{\gamma p} = 0.25 \pm 0.06$,
suggests large absorption corrections, of the order of $60\%$, suppressing
the rate of the studied reaction $\gprho$.
%
%


\begin{thebibliography}{99}
%
\bibitem{DPP}
  N.F. Bali,  G.F. Chew and A. Pignotti, {\em Phys. Rev. Lett.} {\bf 19} (1967) 614; \\
  G.F. Chew and A. Pignotti, {\em Multiperipheral Bootstrap Model}, 
  {\em Phys. Rev.} {\bf 176} (1968) 2112; \\
  E.L. Berger, {\em Phys. Rev.} {\bf 179} (1969) 1567.
%
\bibitem{DHD}
  S.D. Drell and K. Hiida, \PRL{\bf 7} (1961) 199;  R.T. Deck, \PRL{\bf 13} (1964) 169; \\
  L.A. Ponomarev, Sov. J. Part. Nucl. {\bf 7} (1976) 70; \\
  F. Hayot {\em et~al.},  Lett. Nuovo Cim. {\bf 18} (1977) 185; \\
  G. Cohen-Tannoudji, A. Santoro and M. Souza, \NPB {\bf125} (1977) 445; \\
  N.P. Zotov and V.A. Tsarev, Sov. J. Part. Nucl. {\bf 9} (1978) 266.
%
\bibitem{Pompyt}
  P. Bruni and G. Ingelman, {\em Diffractive hard scattering at e p and p anti-p colliders},
  in proceedings of the {\em Europhysics Conference}, C93-07-22, Marseille, France (1993) 595.
%
\bibitem{DiffVM}
  B.~List and A.~Mastroberardino,
  {\em DIFFVM - A Monte Carlo generator for diffractive processes in ep scattering},
  Proc. of the Workshop on Monte Carlo Generators for HERA Physics,
  eds. A.T. Doyle et al., DESY-PROC-1999-02 (1999) 396.
%
\bibitem{VMD}
  J.J. Sakurai, Annals Phys.\  {\bf 11} (1960) 1; \\
  J.J. Sakurai, \PRL{\bf 22} (1969) 981; \\
  T.H. Bauer {\it et al.}, Rev. Mod. Phys. {\bf 50} (1978) 261.
%
\bibitem{OPE}
  J. D. Sullivan,  \PRD{\bf 5} (1972) 1732; \\
  V. Pelosi, ``One-pion exchange and inclusive reactions'',
  Lett. Nuovo Cim.  {\bf 4} (1972) 502;\\
  G. Levman and K. Furutani, ``Virtual pion scattering at HERA'', DESY-95-142 (1995)
%
\bibitem{pi_flux}
 (a) M. Bishari, {\em Pion exchange and inclusive spectra}, \PLB{\bf 38} (1972) 510; \\
 (b) H. Holtmann, A. Szczurek and J. Speth
     \NPA{\bf 596} (1996) 631;
     M.Przybycien, A.Szczurek and G.Ingelman,
     \ZPC{\bf 74} (1997) 509; \\
 (c) B.Kopeliovich, B.Povh and I.Potashnikova,
     \ZPC{\bf 73} (1996) 125; \\
 (d) W. Melnitchouk, J. Speth and A.W.Thomas,
 (e) L. Frankfurt, L. Mankiewicz and M. Strikman,  
     \ZPA{\bf 334} (1989) 343; \\
 (f) N.N. Nikolaev, W.Sch\"afer, A. Szczurek and J. Speth,
     \PRD{\bf 60} (1999) 014004.
%
\bibitem{H1paper}
  V.~Andreev {\it et al.} [H1 Collaboration],
  \EJC{\bf 76} (2016) 41.
%
\bibitem{abs_corr}
  N. Nikolaev, J. Speth and B.G. Zakharov, hep-ph/9708290; \\
  U. D'Alesio and H.J. Pirner, \EJA {\bf 7} (2000) 109 [hep-ph/9806321]; \\
  A.B. Kaidalov {\em et~al.}, \EJC{\bf 47} (2006) 385 [hep-ph/0602215]; \\
  B.Z. Kopeliovich {\em et~al.}, \PRD{\bf 85} (2012) 114025 [arXiv:1205.0067].
%
\bibitem{ZEUS_pi2p}
  S.~Chekanov {\it et al.}  [ZEUS Collaboration],
  \NPB{\bf 637} (2002) 3.
%
\end{thebibliography}
\end{document}